\documentclass[12pt,preprint]{aastex}

\def\f{\frac}

\def\max{{\rm max}}
\def\min{{\rm min}}
\def\sj{{\scriptscriptstyle (j)}}
\def\tot{{{\rm tot}}}
\def\sub{{{\rm sub}}}
\def\obs{{{\rm obs}}}
\def\dep{{{\rm dep}}}
\def\dur{{{\rm dur}}}

\slugcomment{published in ApJ}

\shorttitle{JET STRUCTURE OF GRBs}
\shortauthors{TAKAMI ET AL.}

\begin{document}


\title{PROBING THE STRUCTURE OF GAMMA-RAY BURST JETS WITH THE
STEEP DECAY PHASE OF THEIR EARLY X-RAY AFTERGLOWS}


\author{
Kentaro Takami\altaffilmark{1}, 
Ryo Yamazaki\altaffilmark{1},
Takanori~Sakamoto\altaffilmark{2}, 
and Goro~Sato\altaffilmark{2} 
}

\altaffiltext{1}{
Department of Physics, Hiroshima University, Higashi-Hiroshima,
Hiroshima 739-8526, Japan;
takami@theo.phys.sci.hiroshima-u.ac.jp,
ryo@theo.phys.sci.hiroshima-u.ac.jp.
}
\altaffiltext{2}{
NASA Goddard Space Flight Center, Greenbelt, MD 20771.}


\begin{abstract}
We  show that the jet structure of gamma-ray bursts(GRBs)
can be investigated with the tail emission of the prompt
GRB. The tail emission that we consider is identified as a 
steep decay component of the early X-ray afterglow observed by
the X-Ray Telescope on board {\it Swift}.
Using a Monte Carlo method, we derive for the first time
the distribution of the decay index of the GRB tail emission
for various jet models.
The new definitions of the zero of time and the time interval 
of a fitting region are proposed.
These definitions for fitting the light curve 
lead us to a unique definition of the decay index, 
which is  useful to investigate the structure 
 of the GRB jet.
We find that
if the GRB jet has a core-envelope structure, the predicted distribution
of the decay index of the tail has a wide scatter and
multiple peaks, which cannot be seen for the case of the uniform- and
the Gaussian jet.
Therefore, the decay index distribution gives us information
about the jet structure.
Especially if
 we observe events whose decay index is less than about 2,
 both the uniform- and the Gaussian jet models
 will be disfavored, according to our 
simulation study.

%

\end{abstract}


%
\section{Introduction}
\label{sec:intro}

Gamma-ray burst (GRB) jet structure, that is, the energy distribution
$E(\theta)$ in the ultra-relativistic collimated  outflow,
 is at present not yet fully understood \citep{zhang02}.
There are many jet models proposed in addition to the simplest
uniform-jet model:
the power-law jet model \citep{rossi02,zhang02st}, the
Gaussian jet model \citep{zhang04}, the annular jet model \citep{eichler04},
the multiple emitting subshell model \citep{kumar00multi,nakamura00},
the two-component jet model \citep{berger03}, and so on.
The jet structure may depend on the generation process of the jet
and therefore may provide us important information about the central engine of 
the GRB.
For example, in the collapsar model 
for long GRBs \citep[e.g.,][]{zhang03c,zhang04c},
the jet penetrates into and breaks out of the progenitor star,
resulting in the $E(\theta)\propto\theta^{-2}$ profile \citep{lazzati05}.
For the compact binary merger model for short GRBs,
hydrodynamic simulations have shown that the
resulting jet tends to have a flat core surrounded by the
power-law-like envelope \citep{aloy05}.

In the pre-{\it Swift} era, there were many attempts to
constrain the GRB jet structure.
Thanks to the {\it HETE-2}, statistical properties of long GRBs,
X-ray-rich GRBs, and X-ray flashes were obtained \citep{sakamoto05},
which were thought to constrain the jet models \citep{lamb04}.
These observational results constrain various jet models,
such as the uniform-jet model \citep{yama04a,lamb05,donaghy06},
the multiple subshell model \citep{toma05}, 
the Gaussian jet model \citep{dai05}, and so on.
For BATSE long GRBs, \citet{yone05} derived 
the distribution of the pseudo-opening angle, inferred from
the Ghirlanda \citep{ghir04} and Yonetoku \citep{yone04} relations,
as $f( \theta_j )d \theta_j\propto \theta_j^{-2}d \theta_j $,
which is compatible with that predicted by 
the power-law jet model as discussed in \citet{perna03}
\citep[however, see][]{nakar04}.
%
Afterglow properties are also expected to constrain the jet structure
\citep[e.g.,][]{granot03}; however, energy redistribution 
effects prevent us from reaching a definite conclusion.
Polarization measurements of optical afterglows bring us 
important information \citep{lazzati04}.

In the {\it Swift} era, rapid follow-up observation reveals prompt
GRBs followed by a steep decay phase in the X-ray early afterglow
\citep{tagliaferri06,nousek06,obrien06}.
In the most popular interpretations, the steep
decay component  is the tail emission of the prompt GRB 
(the so called high-latitude emission), i.e., the internal shock origin 
\citep{zhang06,yama06,liang06,dyks05},
although there are some other possibilities
 \citep[e.g.,][]{kobayashi05,pana06,peer06,lazzati06,dado06}.
Then, for the uniform-jet case, 
the predicted decay index is $\alpha=1-\beta$, 
where we use the convention $F_\nu\propto T^{-\alpha}\nu^{1+\beta}$
\citep{kumar00}.
For power-law jet case ($E(\theta)\propto\theta^{-q}$), 
the relation is modified to 
$\alpha=1-\beta+(q/2)$.
However, these simple analytical relations cannot be directly compared
with observations, because they are for the case in which the observer's
line of sight is along the jet axis and because changing the
zero of time, which potentially lies anywhere within the epoch where we see
the bright pulses, substantially alters the early decay slope.

Recently, \citet{yama06} (Y06) investigated the tail emission of the
prompt GRB, finding that the jet structure can be described and
that the global decay slope is not so much affected by the local 
angular inhomogeneity as it is affected by the global energy distribution.
They also argued that the structured jet model is preferable, because
steepening GRB tail breaks appeared in some events.
In this paper,
 we calculate for the first time
 the distribution of the decay index of the
prompt tail emission for various jet models
and find that the derived distributions can be distinguished from each other,
so that the jet structure can be more directly constrained
than previous arguments.
This paper is organized as follows.
We describe our model in \S~\ref{sec:model}.
In \S~\ref{sec:index}, we investigate the distribution of the
decay index of the prompt GRB emission.
Section~\ref{sec:discussion} is devoted to discussions.

%
\section{Tail Part of the Prompt GRB Emission}
\label{sec:model}

We consider the same model as discussed in the previous works 
\citep[Y06;][]{yama04b,toma05a,toma05}.
The whole GRB jet, whose opening half-angle is $\Delta\theta_{\tot}$,
consists of $N_{\tot}$ emitting subshells.
We introduce the spherical coordinate system $(r, \vartheta, \varphi, t)$ in
the central engine frame, where the origin is at the
central engine and $\vartheta=0$ is the axis of the whole jet.
Each emitting subshell departs at time $t_{\dep}^{\sj}$
($0<t_{\dep}^{\sj}<t_{\dur}$, where $j=1,\cdots, N_{\tot}$, and 
$t_{\dur}$ is the active time of the central engine) 
from the central engine in the direction of 
$\vec{n}^{\sj}=(\vartheta^{\sj}, \varphi^{\sj})$,
and emits high-energy photons, generating a single pulse
as observed.
The direction of the observer is denoted by 
$\vec{n}_{\rm obs}=(\vartheta_{\obs}, \varphi_{\obs})$.
The observed flux from the $j$th subshell
is calculated when the following parameters are determined:
the viewing angle of the subshell $\theta_v^{\sj}
=\cos^{-1}(\vec{n}_{\obs}\cdot\vec{n}^{\sj})$,
the angular radius of the emitting shell $\Delta\theta_{\sub}^{\sj}$,
the departure time $t_{\dep}^{\sj}$,
the Lorentz factor $\gamma^{\sj}=(1-\beta_\sj^2)^{-1/2}$, 
the emitting radius $r_0^{\sj}$, 
the low- and high-energy photon indices $\alpha_B^{\sj}$ and $\beta_B^{\sj}$, 
the break frequency in the shell comoving frame ${\nu'_0}^{\sj}$
 \citep{band93},
the normalization constant of the emissivity $A^{\sj}$, 
and the source redshift $z$.
The observer time $T=0$ is chosen as the time of arrival at the 
 observer of a photon emitted at the origin $r=0$ at $t=0$.
Then, at the observer,
the starting and ending times of the $j$th subshell emission
are given by
\begin{eqnarray}
T_{\rm start}^\sj &\sim&
t_{\dep}^\sj+
\f{r_0^\sj}{2c\gamma_\sj^2}\left(1+\gamma_\sj^2{\theta_-^\sj}^2
\right)~~ ,
\label{eq:Tstart}\\
T_{\rm end}^\sj &\sim& 
t_\dep^\sj+
\f{r_0^\sj}{2c\gamma_\sj^2}\left(1+\gamma_\sj^2{\theta_+^\sj}^2
\right) ~~,
\label{eq:Tend}
\end{eqnarray}
where $\theta_+^\sj=\theta_v^\sj+\Delta\theta_\sub^\sj$,
$\theta_-^\sj=\max\{ 0,\theta_v^\sj-\Delta\theta_\sub^\sj \}$, and
we use the formulas $\beta_\sj\sim1-1/2\gamma_\sj^2$
and $\cos\theta\sim1-\theta^2/2$ 
for $\gamma^\sj\gg1$ and $\theta\ll1$, respectively.
The whole light curve from the GRB jet is produced by the superposition
of the subshell emission.

Y06 discussed some kinematical properties of prompt GRBs in
our model and found that each
emitting subshell with $\theta_v^\sj\gg\Delta\theta_\sub^\sj$
produces a single, smooth, long-duration, dim,  and soft pulse,
and that such pulses overlap with each other and
make the tail emission of the prompt GRB.
Local inhomogeneities in the model are almost
averaged during the tail emission phase, and 
the decay index of the tail is determined by
the global jet structure, that is
the mean angular  distribution of the emitting subshell
because in this paper all subshells are assumed to have the same properties
unless otherwise stated.
Therefore, we are essentially studying the tail emission from
the usual continuous jets at once, i.e., from uniform- or power-law jets
with no local inhomogeneity.
In the following, we  study various energy distributions
of the GRB jet through the change of the angular distribution 
of the emitting subshell.
%

%
\section{Decay Index of the Prompt Tail Emission}
\label{sec:index}

In this section, we perform Monte~Carlo simulations 
in order to investigate the jet structure by calculating
 the statistical properties of the decay index of the tail emission.
For a fixed-jet model, we randomly generate $10^4$ observers with their
line of sights (LOSs) 
$\vec{n}_{\rm obs}=(\vartheta_{\obs},\varphi_{\obs})$.
For each LOS,  the light curve, $F(T)$ 
of the prompt GRB tail in the 15--25~keV band is calculated,
and the decay index is determined.
The adopted observation band is the low-energy end of the 
Burst Alert Telescope(BAT) detector
and near the high-energy end of the 
X-Ray Telescope(XRT) on {\it Swift}.
Hence, one can observationally obtain continuous light curves, 
beginning with the
prompt GRB phase to  the subsequent early afterglow phase
\citep{sakamoto07}, so that it is
convenient for us to compare theoretical results with
 observations.
However, our actual calculations have shown that our
conclusion 
is not qualitatively altered,
even if the
observation band is changed, for example,
to 0.5--10~keV, as usually considered for other references.

\begin{figure}
\epsscale{.50}
\plotone{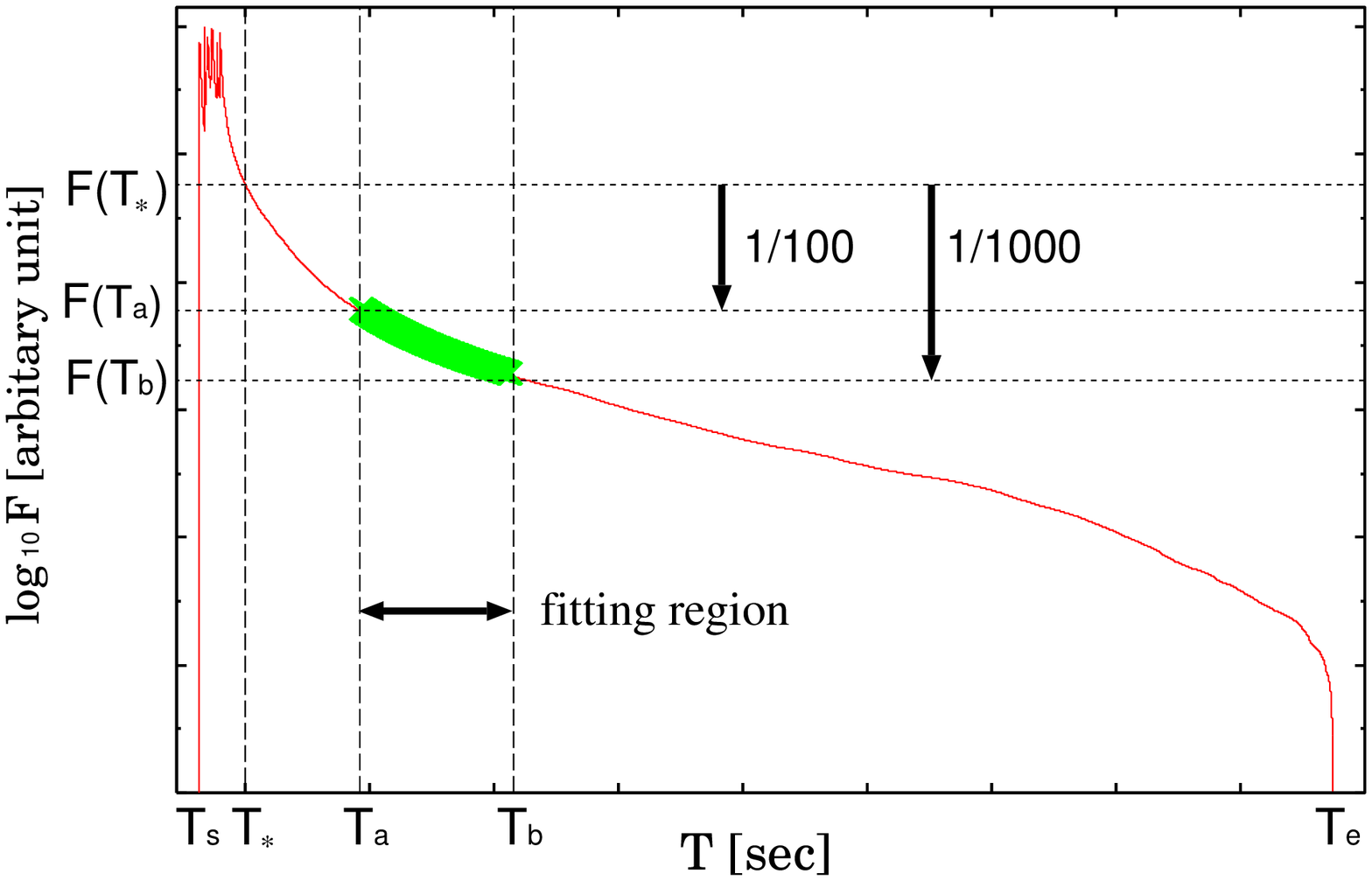}
\caption{
Example of how the decay index $\alpha$ is determined by the
calculated light curve $F(T)$.
The start and end time of the burst are denoted by $T_{\rm s}$
and $T_{\rm e}$, respectively.
The time $T_{*}$ is determined by Eq.~(\ref{eq:index}).
The decay index $\alpha$ is determined by fitting
$F(T)\propto(T-T_*)^{-\alpha}$ in the time interval
$[T_{a},T_{b}]$.
}
\label{fig1}
\end{figure}

For each light curve, the decay index is calculated by
 fitting $F(T)$ with a single power-law form, 
$\propto(T-T_*)^{-\alpha}$, as in the following
(see Fig.~\ref{fig1}).
The decay index $\alpha$ depends on the choice of $T_*$
\citep{zhang06,yama06}\footnote{
Recently, \citet{kobayashi06} 
have discussed the way to choose the time zero.
According to their arguments, the time zero is near the rising epoch
of the last bright pulse in the prompt GRB phase.
}.
%
Let $T_{\rm s}$ and $T_{\rm e}$ be the start and end time, 
respectively, of the prompt GRB, i.e., 
\begin{eqnarray}
T_{\rm s}&=&\min\{T_{\rm start}^\sj\} \\
T_{\rm e}&=&\max\{T_{\rm end}^\sj\} ~~.
\end{eqnarray}
Then, we take $T_*$ as the time until
$99\%$ of the total fluence, which is defined by 
$S_{\rm total}=\int_{T_{\rm s}}^{T_{\rm e}} F(T')~dT'$,
is radiated, that is,
%
%
%
\begin{equation}
\int_{T_{\rm s}}^{T_*} F(T')~dT'= 0.99\,
S_{\rm total}  ~~.
\label{eq:index}
\end{equation}
Then, the prompt GRB is in the main emission phase for
$T<T_*$, while it is in the tail emission phase for $T>T_*$.
The time interval 
$[T_a,T_b]$,
in which  the decay index $\alpha$ is determined
assuming the form $F(T)\propto(T-T_*)^{-\alpha}$,
is taken to satisfy
%
%
\begin{equation}
F(T_{a,b}) = q_{a,b}F(T_{*})~~,
\label{eq:Tab}
\end{equation}
where we adopt $q_a=1\times10^{-2}$ and $q_b=1\times10^{-3}$, 
unless otherwise stated.
We find that in this epoch the assumed fitting form
gives a well approximation.

At first, we consider the uniform-jet case, in which
the number of subshells per unit solid angle is approximately
given by $dN/d\Omega=N_\tot/(\pi\Delta\theta_\tot^2)$
for $\vartheta<\Delta\theta_\tot$, where $\Delta\theta_\tot=0.25$~rad is
adopted. 
The departure time of each subshell $t_{\dep}^{\sj}$ is assumed to be
homogeneously random between $t=0$ and $t=t_{\dur}=20$~sec.
The central engine is assumed to produce $N_{\tot}=1000$ subshells.
In this section, we assume that all subshells have the same 
values of the following fiducial parameters: $\Delta\theta_{\sub} = 0.02$~rad,
$\gamma=100$, $r_0 = 6.0 \times 10^{14}$~cm, $\alpha_B = -1.0$,
$\beta_B = -2.3$, ${h\nu'_0}=5$~keV,
and $A={\rm constant}$.
%
%
Our assumption of constant $A$ is justified as follows.
%
Note that
the case in which $N$ subshells that have the same brightness $A$ are launched
into the same direction, but a different departure time, is equivalent to
the case of one subshell emission with the brightness of $NA$.
This is because in the tail emission phase, the second terms in the r.h.s.
of Eqs.~(1) and (2) dominate the first terms, so that the 
time difference effect, which arises from the difference of $t_{\rm
dep}$ for each subshell, can be obscured.
Hence, giving the angular distribution of the emission energy
is equivalent to giving the angular distribution of the subshells with
constant $A$.
Also, Y06 showed that
to obtain a smooth, monotonic tail 
emission as observed by {\it Swift}, 
the subshell properties, $h{\nu'_0}^{\sj}$
and/or $A^{\sj}$,
cannot have wide scatter in the GRB jet. 
Therefore, we can expect, at least as the zeroth-order
approximation, that the subshells have the same properties.
%
%

\begin{figure}
\epsscale{1.0}
\plottwo{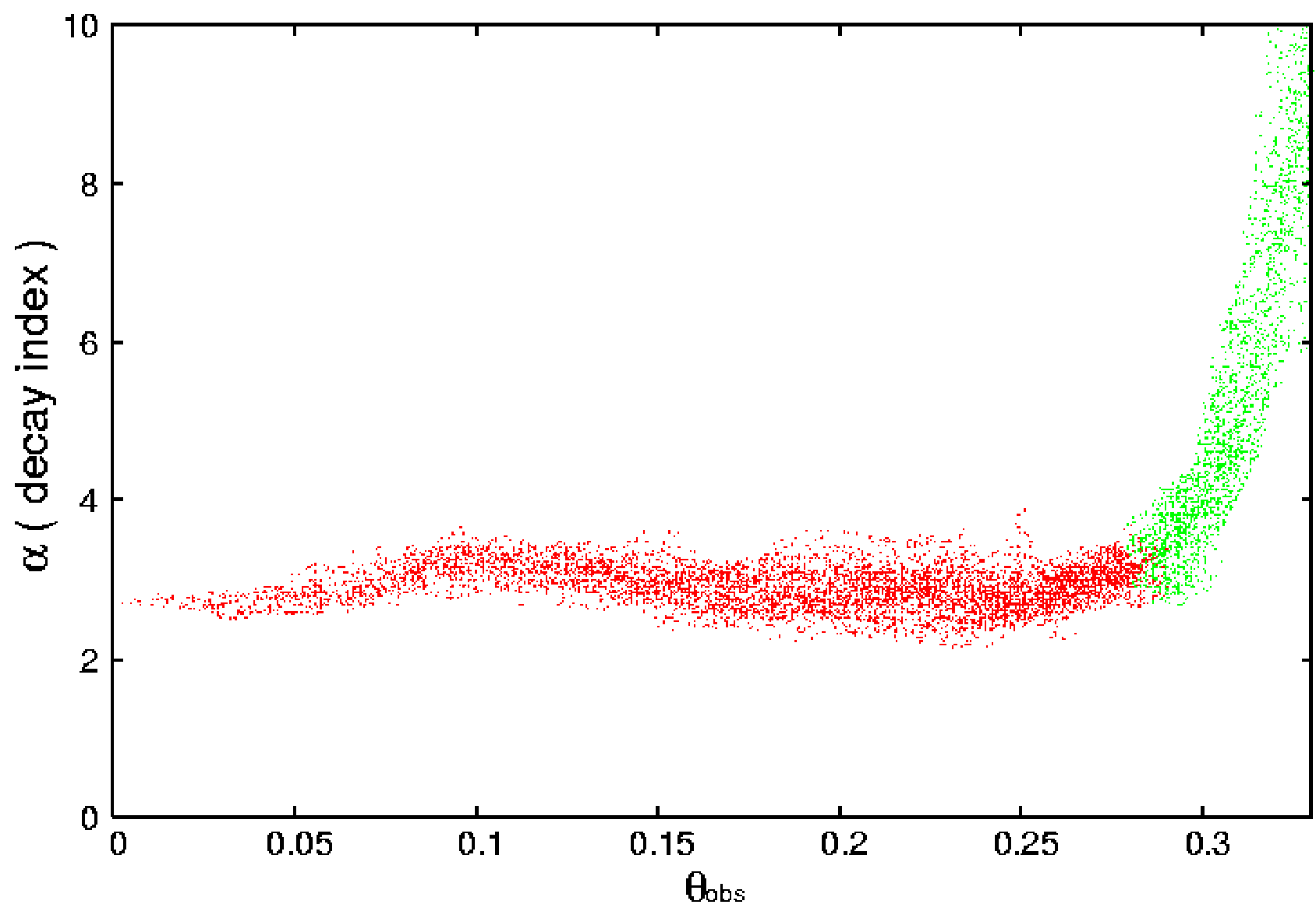}{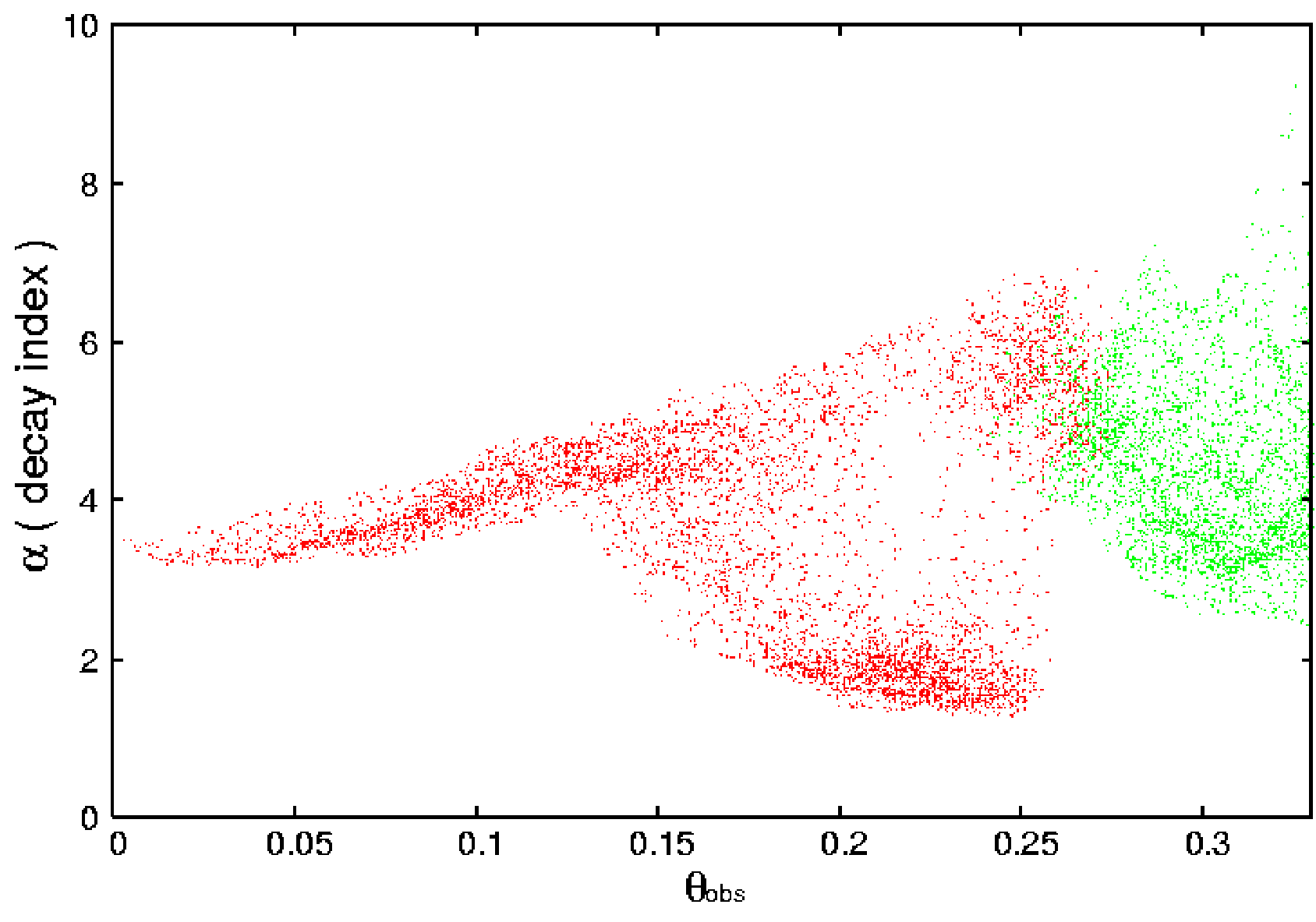}
\caption{
Decay index $\alpha$ as a function of $\vartheta_{\rm obs}$,
the angle between the whole jet axis and the observers' lines of
sight. Red and green points represent events
 whose peak fluxes are larger and smaller than 
$10^{-4}$ times the largest one in all simulated events, 
respectively.
Left and right  panels are for the uniform- and power-law jet
cases, respectively.
}
\label{fig2}

\end{figure}

\begin{figure}
\epsscale{1.0}
\plottwo{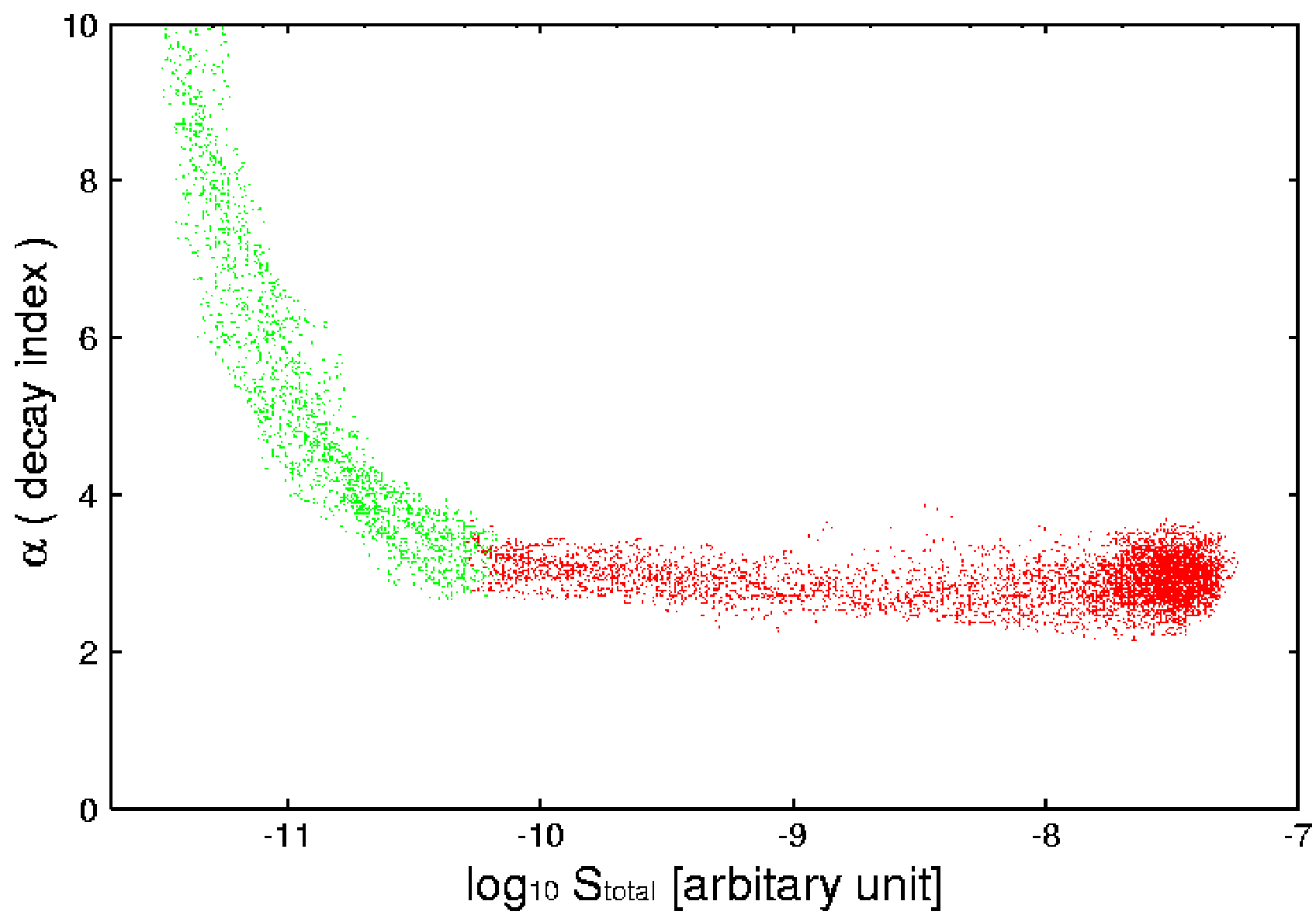}{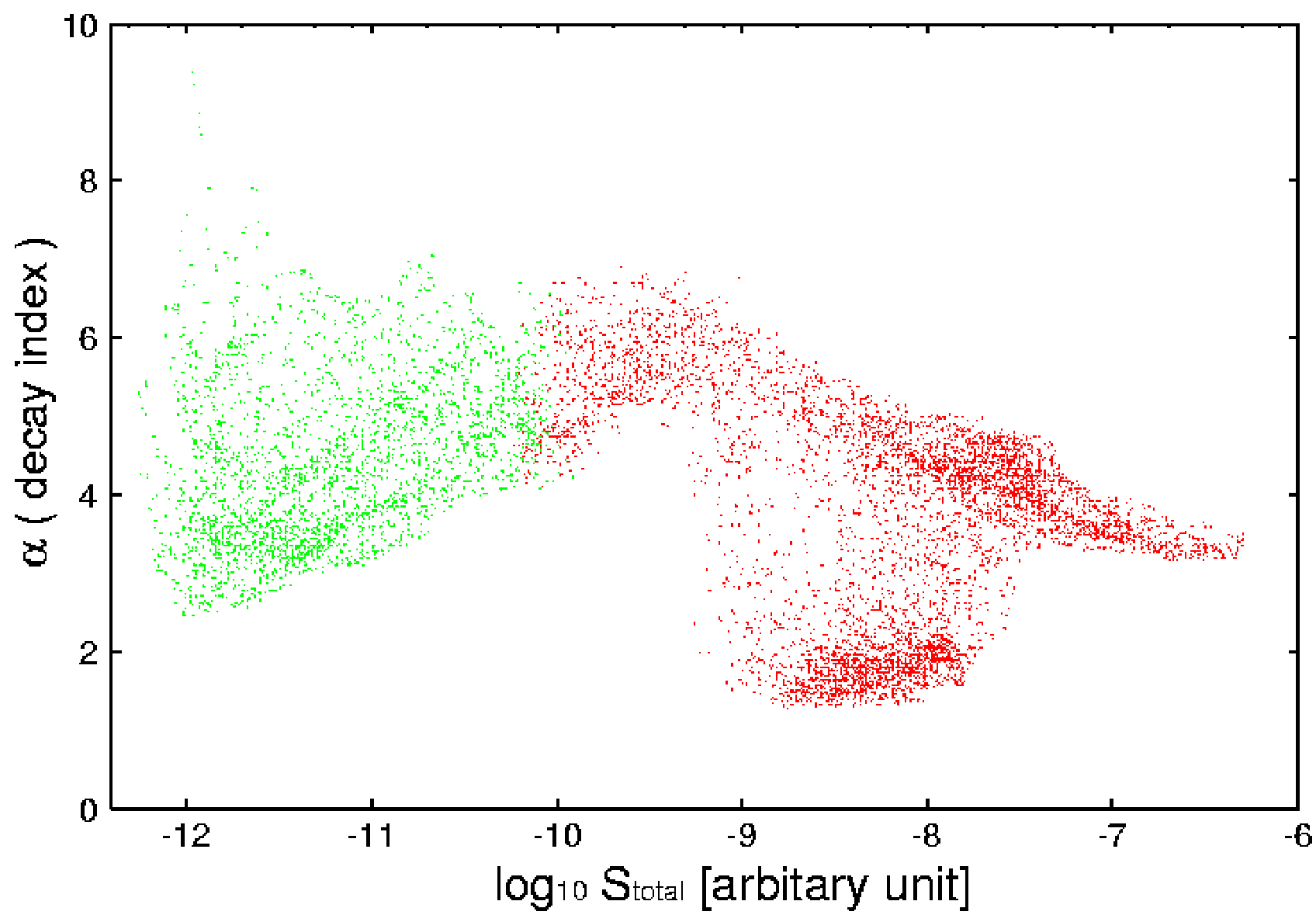}
\caption{
Decay index $\alpha$ as a function of the total fluence 
$S_{\rm total}$, the sum of the fluxes in the time interval 
$[T_{\rm s},T_{\rm e}]$. 
Red and green points represent events whose peak fluxes are larger and
 smaller than $10^{-4}$ times the largest one in all simulated
 events, respectively.
Left and right panels are for the uniform- and power-law jet cases,
 respectively. 
}
\label{add_fig}
\end{figure}

The left panel of Fig.~\ref{fig2} shows the  decay
index $\alpha$ as a function of $\vartheta_{\rm obs}$. 
For $\vartheta_{\obs}\lesssim\Delta\theta_{\tot}$
(on-axis case),
$\alpha$ clusters around $\sim3$.
On the other hand,
when $\vartheta_{\obs}\gtrsim\Delta\theta_{\tot}$
(off-axis case),
$\alpha$ rapidly increases with $\vartheta_{\rm obs}$. 
The reason is as follows.
If all subshells are seen sideways (that is,
$\theta_v^\sj\gg\Delta\theta_\sub^\sj$ for all $j$),
the bright pulses in the main emission phase
followed by the tail emission disappear
 because of the relativistic beaming effect, resulting
in a smaller flux contrast between the main  emission
phase and the tail emission phase compared with the
on-axis case.
Then $T_*$ becomes larger. Furthermore,
in the off-axis case, the tail emission decays more slowly
($|dF/dT|$ is smaller) than in the on-axis case.
Then both $T_a-T_{*}$ and $T_b-T_{*}$ 
 are larger  for the off-axis case than for the on-axis case.
As can be seen in Fig.~3 of \citet{zhang06},
the emission seems to decay rapidly, so that
the decay index $\alpha$ becomes large.
%
%
%
The left panel of Fig.~\ref{add_fig}
shows $\alpha$ as a function of the total fluence $S_{\rm total}$ which
is the sum of the fluxes in the time interval, $[T_{\rm s},T_{\rm e}]$. 
In Fig.~\ref{add_fig}, both  $\alpha$ and $S_{\rm total}$ are 
determined observationally, so that our theoretical calculation can 
be directly compared with the observation.
%
%
%

%
%
A more realistic model is  the Gaussian jet model, 
in which the number of
subshells per unit solid angle is approximately given by 
$dN/d\Omega = C \exp(-\vartheta ^2 / 2\vartheta _c ^2) $
for $0\leqq\vartheta\leqq\Delta\theta_\tot$, where
$C=N_\tot / 2\pi\vartheta_c^2[1-\exp(-\Delta\theta_\tot ^2 /
2\vartheta_c ^2 )]$
is the normalization constant.
We find only a slight difference between the results for the 
uniform- and the Gaussian jet models.
%
%
Therefore, we do not show the results for  the Gaussian jet case
in this paper.
%
%

Next, we consider the power-law distribution. 
In this  case, the number of subshells
per unit solid angle is approximately given by 
$dN/d\Omega=C[1+(\vartheta/\vartheta _c)^2]^{-1}$ 
for $0\leqq\vartheta\leqq\Delta\theta_\tot$, i.e., 
$dN/d\Omega \thickapprox C$ for $0\leqq\vartheta\ll\vartheta_c$ and
$dN/d\Omega\thickapprox C(\vartheta/\vartheta_c)^{-2}$ for
$\vartheta_c\ll\vartheta\leqq\Delta\theta_\tot$,
where 
$C=(N_\tot/\pi\vartheta_c^2)[\ln(1+(\Delta\theta_\tot/\vartheta_c)^2)]^{-1}$
is the normalization constant and we adopt
$\vartheta_c=0.02$~rad and $\Delta\theta_\tot=0.25$~rad.
The other parameters are the same as for the uniform-jet case.

\begin{figure}
\epsscale{1.0}
\plotone{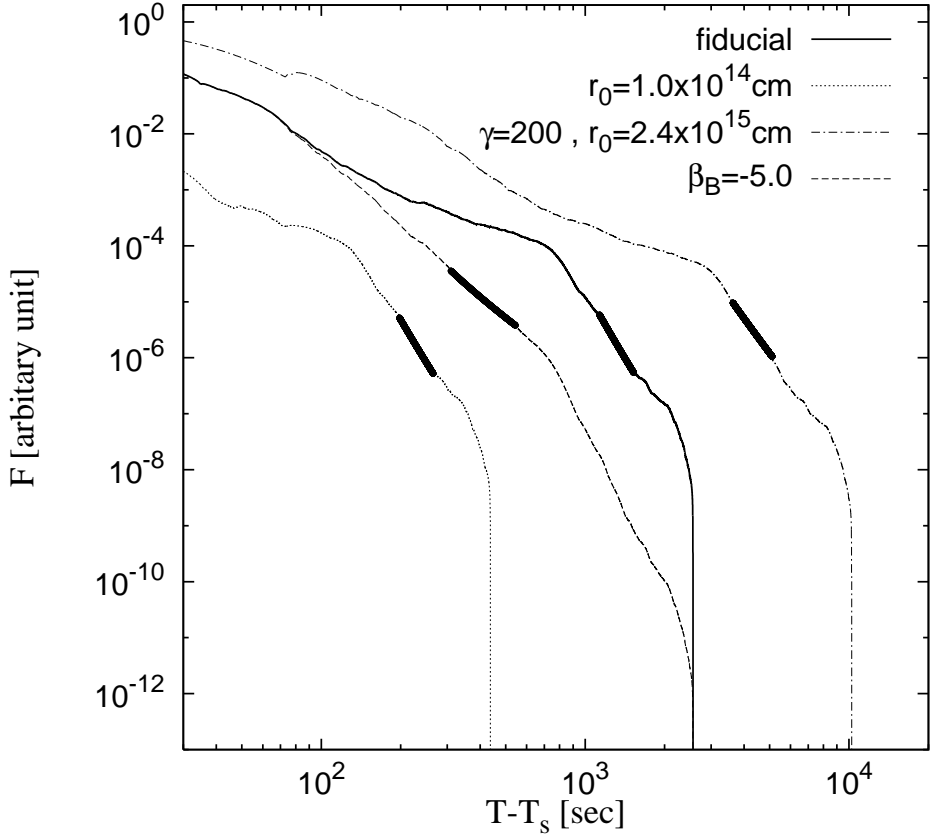}
\caption{
Examples of light curves of the prompt tail emission
in the 15--25~keV band
for the power-law jet case
and $\vartheta_{\rm obs}>\vartheta_c$
($\vartheta_{\rm obs}=0.27$~rad and $\vartheta_c=0.02$~rad).
The solid line shows the fiducial parameters.
A bump caused by the core emission can be seen at
$T-T_s\sim7.5\times10^2$~s.
The dotted, dot-dashed, and dashed lines are for
$r_0=1.0\times10^{14}$~cm; 
$r_0=2.4\times10^{15}~cm$ and $\gamma=200$; 
and $\beta_B=-5$, respectively,
with other parameters being fiducial.
Time intervals $[T_a,T_b]$ for each case are denoted
by the thick solid lines.
The flux is normalized by the peak value.
}
\label{fig_lightcurve}
\end{figure}

As can be seen in the right panels of Figs.~\ref{fig2} and \ref{add_fig},
both the $\vartheta_\obs$--$\alpha$ 
and $S_{\rm total}$--$\alpha$ diagrams
are complicated
compared with the uniform-jet case.
When $\vartheta_{\obs}\lesssim\vartheta_c$, 
the observer's LOS is near the whole jet axis.
Compared with the uniform-jet case,
 $\alpha$ is larger,  because the power-law jet is dimmer 
in the outer region, i.e., emitting subshells are sparsely
distributed near the periphery of the whole jet
(see also the solid lines of Figs.~1 and 3 of Y06).
If $\vartheta_{\obs}\gg\vartheta_c$,
the scatter of $\alpha$ is large.
Some bursts have an especially small $\alpha$ of around 2.
This comes from the fact that the power-law jet has a core region
($0<\vartheta\lesssim\vartheta_c$), where emitting subshells
densely distributed compared with the outer region.
The core generates  the light-curve break in the tail emission phase,
as can be seen in Fig.~\ref{fig_lightcurve} (Y06).
In the epoch before the photons emitted by the core arrive at the observer 
(e.g., $T-T_{\rm s}\lesssim7.5\times10^2$~s for the solid line in Fig.~\ref{fig_lightcurve}), 
the number of subshells that contribute to the
flux at time $T$, $N_\sub(T)$, increases with $T$ more rapidly 
than for the uniform-jet case.
Then, the light curve shows a gradual decay.
If the fitting region $[T_a,\,T_b]$ lies in this epoch,
the decay index $\alpha$ is around 2.
In the epoch after the photons arising from the core are observed
(e.g., $T-T_{\rm s}\gtrsim7.5\times10^2$~s for the solid line 
in Fig.~\ref{fig_lightcurve}),
the subshell emission with 
$\theta_v^\sj\gtrsim\vartheta_\obs+\vartheta_c$ is observed. 
Then $N_\sub(T)$ rapidly decreases with $T$, and
the observed flux suddenly drops. 
If the interval $[T_a,\,T_b]$ lies in this epoch,
the decay index becomes larger than 4.
%
%
%
%

\begin{figure}
\epsscale{1.0}
\plotone{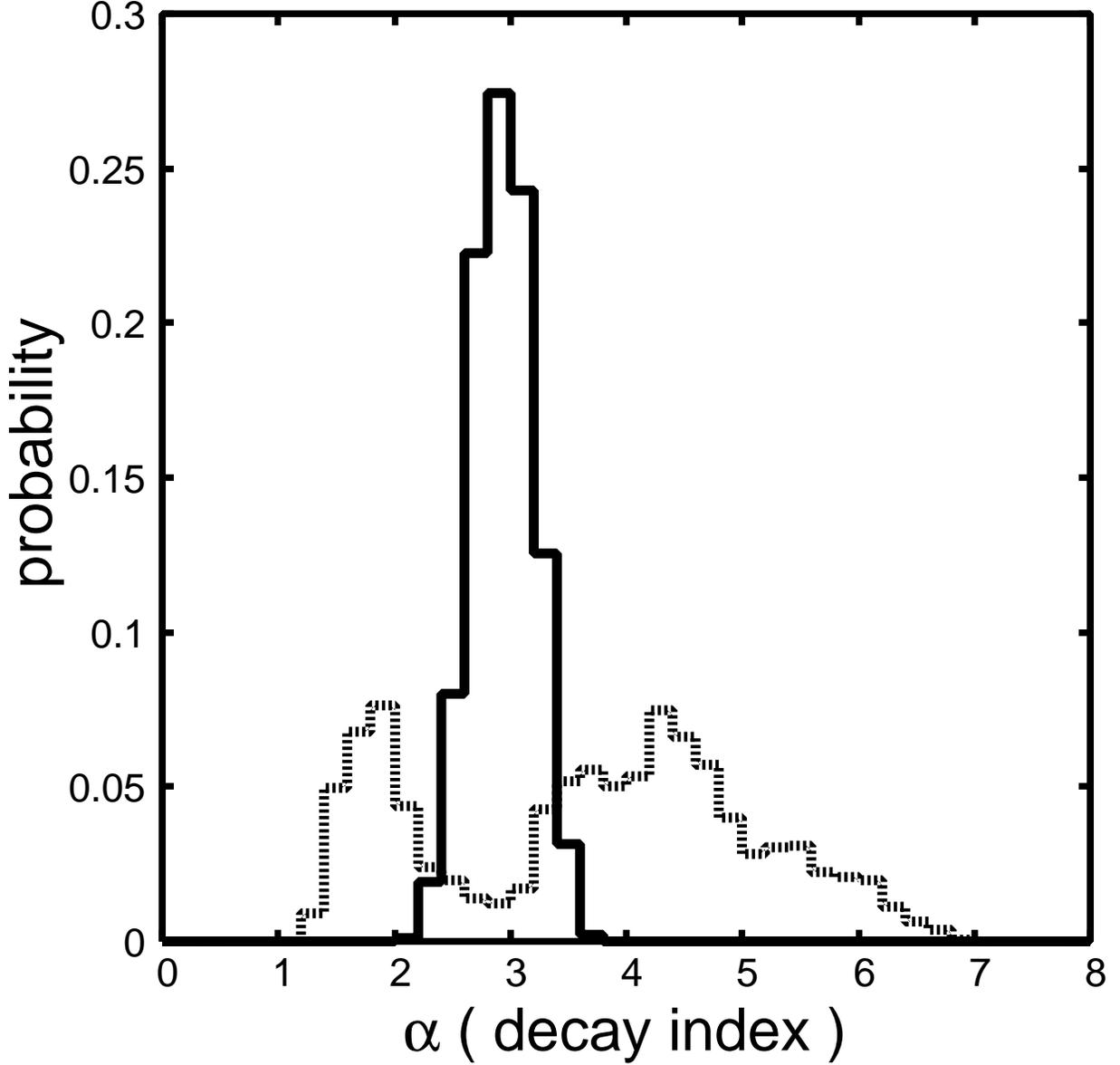}
\caption{
Distributions of the decay index $\alpha$ for uniform-jet 
($dN/d\Omega=const.$;{\it solid line}) and power-law jet 
($dN/d\Omega\propto[1+(\vartheta/\vartheta _c)^2]^{-1}$;{\it dotted line})
models, respectively.
We assume that all subshells have the same 
values of the following fiducial parameters: 
$\Delta\theta_{\sub} = 0.02$~rad,
$\gamma=100$, $r_0 = 6.0 \times 10^{14}$~cm, $\alpha_B = -1.0$,
$\beta_B = -2.3$, ${h\nu'_0}=5$~keV,
and $A={\rm constant}$.
We consider events whose peak fluxes are larger  than 
$10^{-4}$ times the largest one in all simulated events
(red points in Fig.~\ref{fig2}).
}
\label{fig3}
\end{figure}

To compare the two cases considered above more clearly,
 we derive the distribution of the decay index $\alpha$. 
Here we consider the events whose peak fluxes are larger than 
$10^{-4}$ times of the largest one in all simulated events, 
because the events with small peak fluxes are not observed.
Fig.~\ref{fig3} shows the result.
For the uniform-jet case ({\it solid line}),
$\alpha$ clusters around 3, while 
 for the power-law jet case ({\it dotted line}),
the distribution
is broad ($1\lesssim\alpha\lesssim7$) and has multiple
peaks.

\begin{figure}
\epsscale{1.0}
\plotone{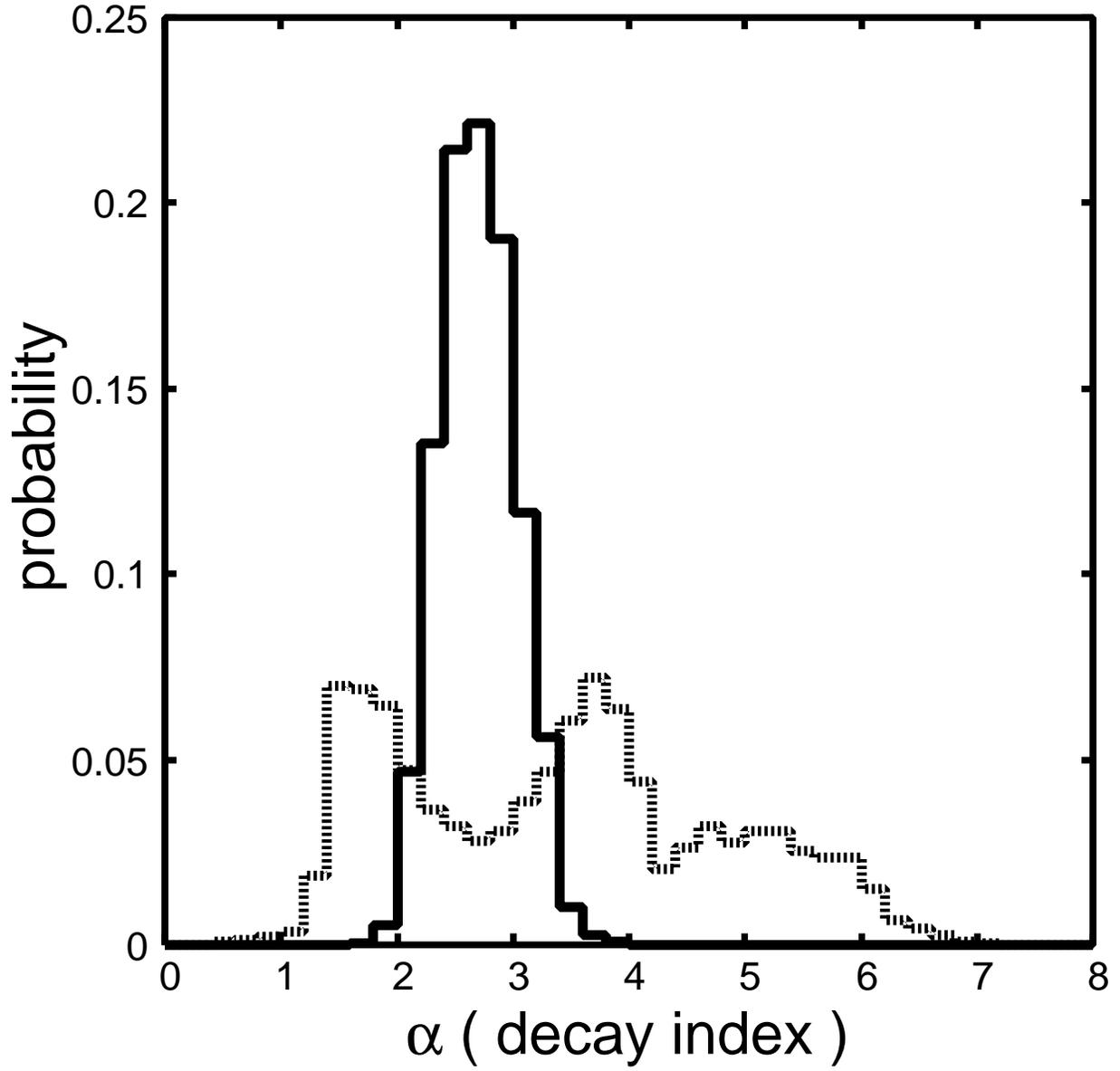}
\caption{
Same as Fig.~\ref{fig3}, but for $r_0=1.0\times10^{14}$~cm.
}
\label{fig4}
\end{figure}

So far, we have considered the fiducial parameters.
In the following, we discuss the dependence on
parameters, $r_0$, $\gamma$, $\beta_B$, $t_{\rm dur}$, and
$\Delta\theta_\tot$
(It is found that the $\alpha$-distribution hardly depends on
the value of $\alpha_B$, $\Delta\theta_\sub$, and $\nu'_0$ within
reasonable parameter ranges).
At first, we consider the case in which $r_0=1.0\times 10^{14}$~cm 
is adopted, with other parameters being fiducial.
Fig.~\ref{fig4} shows the result.
The shape of the $\alpha$-distribution is almost the same as that for the
fiducial parameters, in both the uniform- and the power-law jet cases.
This comes from the fact that in a tail emission phase,
the light curve for a given $r_0$ 
is approximately written as $F(T;r_0)\thickapprox g(cT/r_0)$, where a 
function $g$ determines the light-curve shape of the tail emission for 
other given parameters.
Then, the light curves in the case of $r_0=r_{0,1}$ and $r_0=r_{0,2}$, 
namely, $F(T;r_{0,1})$ and $F(T;r_{0,2})$, 
satisfy the relation
$F(T;r_{0,2})\thickapprox F((r_{0,1}/r_{0,2})T;r_{0,1})$.
This can be seen, for example, by comparing the solid line with the 
dotted one in Fig.~\ref{fig_lightcurve}.
Hence, $T_*$, $T_a$, and $T_b$ are  approximately
proportional to $r_0$; in this simple scaling,
one can easily find that $\alpha$ remains unchanged
for different values of $r_0$.

\begin{figure}
\epsscale{1.0}
\plotone{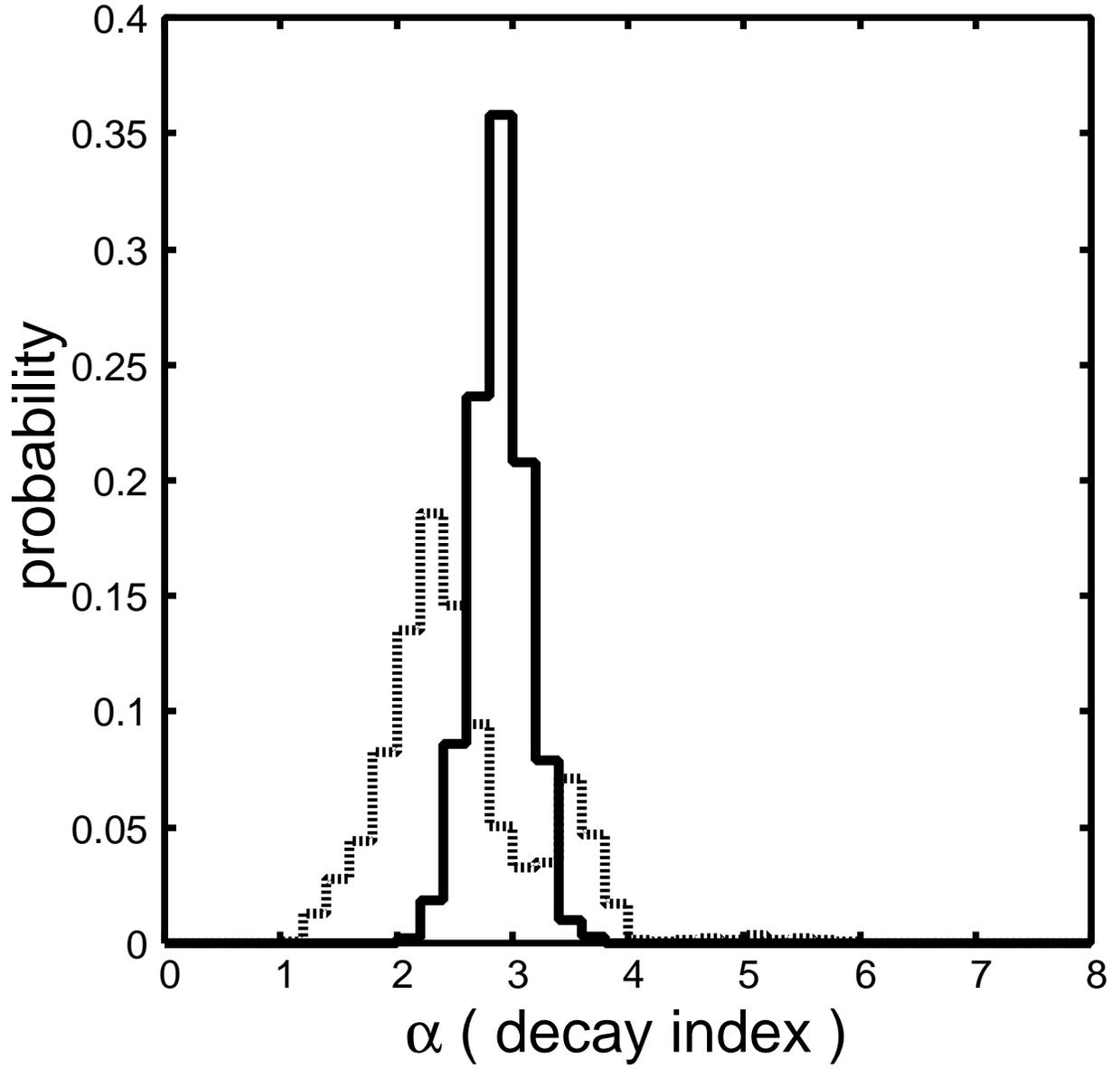}
\caption{
Same as Fig.~\ref{fig3}, but for $\gamma=200$ and 
$r_0=2.4\times 10^{15}$~cm.
}
\label{fig5}
\end{figure}

Second, we consider the case of $\gamma=200$ and
$r_0=2.4\times 10^{15}$~cm, with other parameters being fiducial.
In this case, the angular spreading timescale ($\propto r_0/\gamma^2$)
is the same as in the fiducial case, so that the tail emissions still show
smooth light curves, although the whole emission ends later, according to
the scaling $T_{\rm e}\propto r_0$ 
(see the dot-dashed line in Fig.~\ref{fig_lightcurve}).
Fig.~\ref{fig5} shows the result.
For large $\gamma$, the relativistic beaming effect is more significant,
 so that the events in $\vartheta_{\obs}\gtrsim \Delta \theta_{\tot}$,
which cause large $\alpha$, are dim compared with the small-$\gamma$ case. 
Such events cannot be observed. 
For the power-law jet case, therefore,
the number of large-$\alpha$ events becomes small,
although  the distribution is still
broad ($1\lesssim \alpha \lesssim 4$) and has two peaks.
On the other hand, for the uniform-jet case, 
the distribution of the decay index $\alpha$ is
almost the same as for the fiducial parameter set,
because the value of the decay index $\alpha$ in the case of
$\vartheta_{\obs} \lesssim \Delta \theta_{\tot}$ is almost 
 the same as that in the case of
 $\vartheta_{\obs}\gtrsim \Delta \theta_{\tot}$.

\begin{figure}
\epsscale{1.0}
\plotone{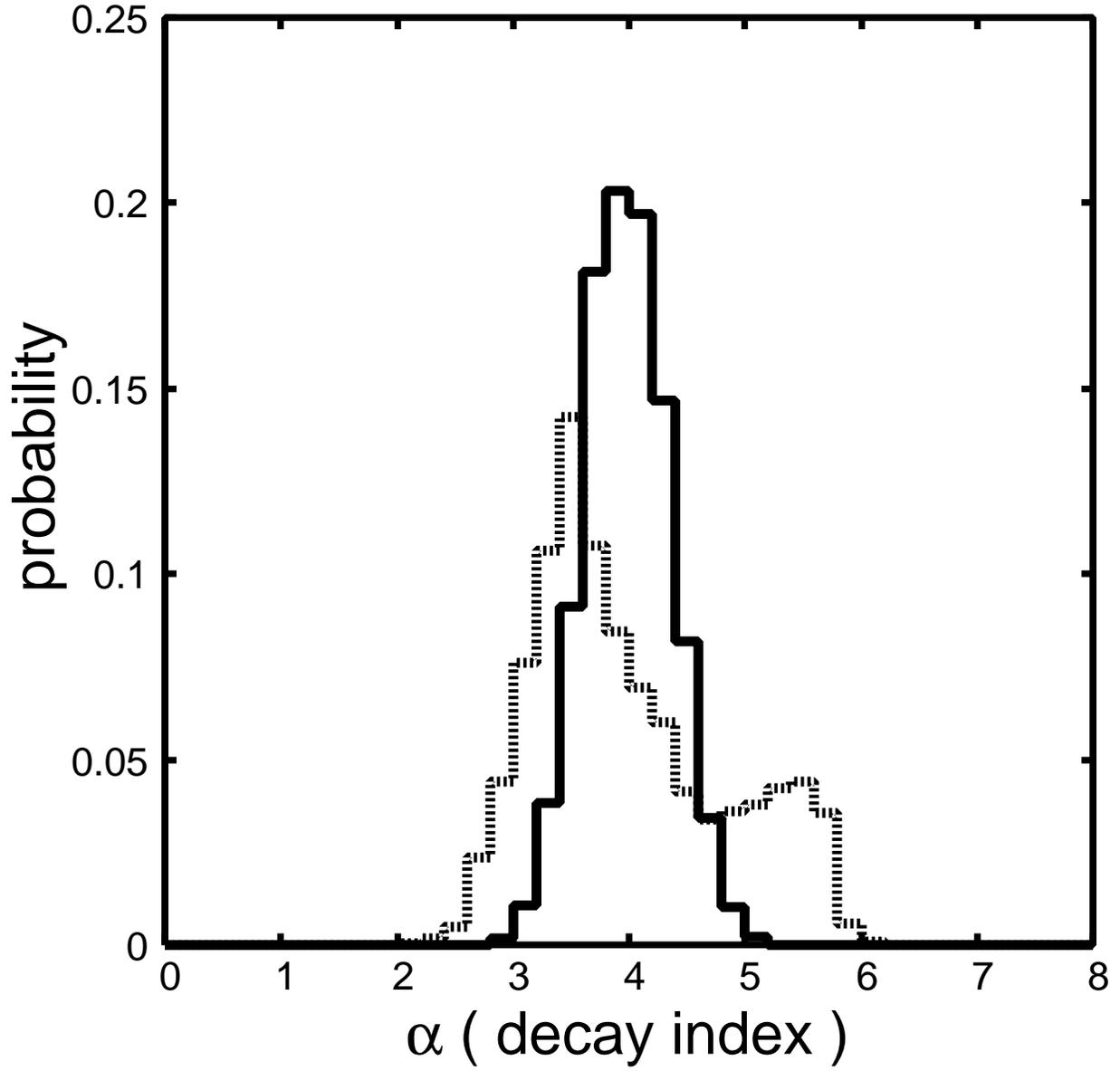}
\caption{
Same as Fig.~\ref{fig3}, but for $\beta_B=-5.0$.
}
\label{fig6}
\end{figure}

Third, we change the value of the high-energy photon index $\beta_B$ 
from $-2.3$ to $-5.0$, with other parameters being fiducial.
Fig.~\ref{fig6} shows the result.
For the uniform-jet case,
 the mean value is $\langle\alpha\rangle\sim4$, while 
$\langle\alpha\rangle\sim3$ for the 
fiducial parameters, so that the decay index defined in this paper
does not obey the well-known formula $\alpha=1-\beta_B$
\citep{kumar00}.
For the power-law jet case, 
the whole distribution shifts toward the higher value,
and the ratio of the two peaks changes.
In the tail emission phase, the spectral peak energy $E_{\rm peak}$
is below 15~keV (see also Y06), so that the steeper the spectral slope
of the high-energy side of the Band function,
the more rapidly the emission decays, resulting in the
dimmer tail emission (see the dashed line in Fig.~\ref{fig_lightcurve}).
Then, the fitting region $[T_a,T_b]$ shifts toward earlier epochs,
because $T_*$ becomes small.
Therefore, the number of events with small $\alpha$ increases,
and the number of events with large $\alpha$ decreases.
Furthermore,
we  comment on the  case in which $\beta_B$ is varied for each event 
in order to more directly compare with the observation.
Here we randomly distribute $\beta_B$ according to the 
Gaussian distribution with 
a mean of $-2.3$ and a variance of $0.4$.
It is found that the results are not qualitatively changed.

\begin{figure}
\epsscale{1.0}
\plotone{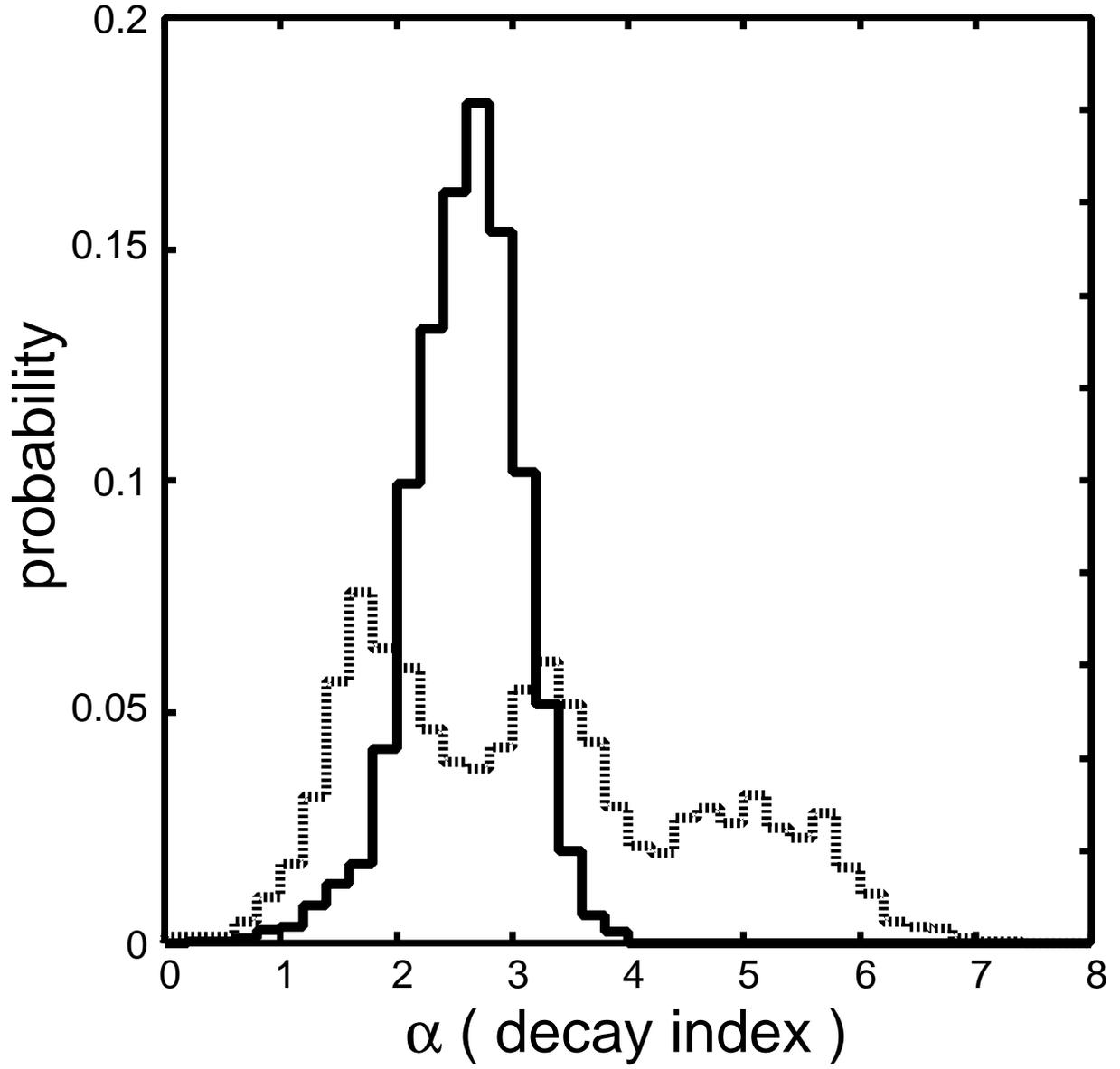}
\caption{
Same as Fig.~\ref{fig3}, but for $t_\dur=200$~sec.
}
\label{fig7}
\end{figure}

Next, we change the value of the duration time $t_\dur$  
from $20$~sec to $200$~sec,
with other parameters being fiducial.
The epoch of the bright pulses in the main emission phase becomes longer
 than that in $t_\dur=20$~sec. 
However, the behavior of the tail emission does not depend on $t_\dur$ 
very much (see Fig.~\ref{fig7}).
Therefore, the distribution of the decay index $\alpha$ is almost the same as
that for the fiducial parameters for both the uniform-jet case 
and the power-law jet case.
Even if we consider the case in which $t_{\rm dur}$ is randomly
distributed for each event
according to the lognormal distribution
with an average of $\log(20~{\rm s})$ and a logarithmic
variance of $0.6$, the results are not significantly changed.

\begin{figure}
\epsscale{1.0}
\plotone{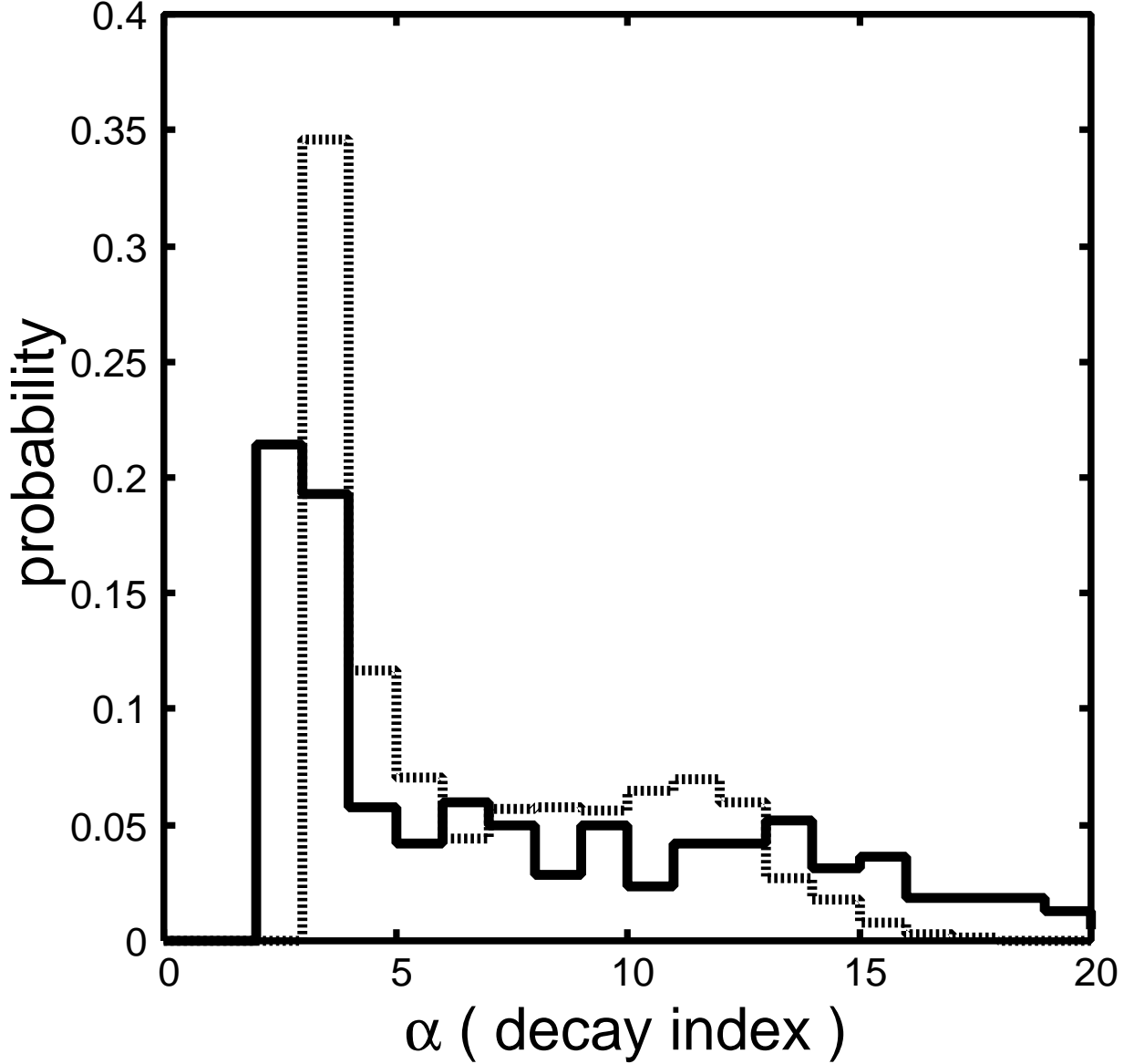}
\caption{
Distribution of the decay index $\alpha$ for the uniform-jet profile.
The dotted line is for $\Delta\theta_\tot=0.1$~rad with other
fiducial parameters.
The solid line is for the variable-$\Delta\theta_\tot$ case,
in which we generate events whose $\Delta\theta_\tot$
distributes as $f_{\Delta\theta_\tot}d(\Delta\theta_\tot)
\propto\Delta\theta_\tot{}^{-2}d(\Delta\theta_\tot)$
($0.05\lesssim\Delta\theta_\tot\lesssim0.4$),
and for given $\Delta\theta_\tot$,
the quantities $\nu'_0$ and $A$ are determined by
$h\nu'_0=(\Delta\theta_\tot/0.13)^{-3.6}$~keV
and $A\propto(\Delta\theta_\tot)^{-7.3}$,
respectively.
Other parameters are fiducial.
}
\label{fig9}
\end{figure}

Finally, we discuss the dependence on $\Delta\theta_\tot$.
Only the uniform-jet case is considered, because 
the structured jet is usually quasi-universal and because
we focus our attention on the behavior of the
uniform-jet model.
The dotted line in Fig.~\ref{fig9} shows the result for
constant $\Delta\theta_\tot=0.1$~rad with other parameters
being fiducial.
We can see many  events with large $\alpha$. 
The large $\alpha$ is observed because for 
small $\Delta \theta_{\rm tot}$,
although the off-axis events 
(i.e., $\Delta \theta_{\rm tot} \lesssim \vartheta_{\rm obs}$)
are still dim because of the relativistic beaming effect, 
a  fraction of such events survives
the flux threshold condition and are observable.
Such events  have large $\alpha\gtrsim5$
(see the 4th paragraph of this section, which explains
the left panel of Fig.~\ref{fig2}).
This does not occur in the large-$\Delta \theta_{\rm tot}$ case.
However, we still find in this case that there are no events 
with $\alpha\lesssim2$.
%
%
We consider another case in which
 $\Delta\theta_\tot$ is variable.
Here we generate events whose $\Delta\theta_\tot$
distributes as $f_{\Delta\theta_\tot}d(\Delta\theta_\tot)
\propto\Delta\theta_\tot{}^{-2}d(\Delta\theta_\tot)$
($0.05\lesssim\Delta\theta_\tot\lesssim0.4$).
Then for a given $\Delta\theta_\tot$,
the quantities $\nu'_0$ and $A$ are determined by
$h\nu'_0=(\Delta\theta_\tot/0.13)^{-3.6}$~keV
and $A\propto(\Delta\theta_\tot)^{-7.3}$,
respectively.
Other parameters are fiducial.
If the model parameters are chosen in this way,
the Amati and Ghirlanda relations 
\citep{amati02,ghir04} are satisfied,
and the event rates of long GRBs, X-ray-rich GRBs and
X-ray flashes become similar \citep{donaghy06}.
The solid line in Fig.~\ref{fig9} shows the result.
Again we find that there are
no events with $\alpha\lesssim2$.

In summary,
when we adopt model parameters within reasonable ranges, 
the decay index becomes larger than $\sim2$
for the uniform- and the Gaussian jet cases,
while a significant fraction of events with $\alpha\lesssim2$
is expected for the power-law jet case.
Therefore, if a non-negligible number of events with
$\alpha\lesssim2$  are observed, both the
uniform- and the Gaussian jet models will be disfavored.
Furthermore, if we observationally derive the 
$\alpha$-distribution, the structure of GRB jets
will be more precisely determined.

%
\section{Discussion}
\label{sec:discussion}

We have calculated the distribution of the decay
index, $\alpha$, for the uniform-, Gaussian, and the power-law jet cases.
For the uniform-jet case, $\alpha$ becomes larger than $\sim2$,
and its distribution has a single peak.  
The Gaussian jet model predicts  almost the same results as
the uniform-jet model.
On the other hand,
 for the power-law jet case,
$\alpha$ ranges between $\sim1$ and $\sim7$,
and its distribution has multiple peaks.
Therefore, we can determine the jet structure of GRBs by analyzing
a lot of early X-ray data showing a steep decay component
that is identified as a prompt GRB tail emission. 
However, one of the big challenges in the {\it Swift} data for calculating 
the decay index in our definition is to derive the composite 
light curve of BAT and XRT.  Since the observed energy bands of 
BAT and XRT do not overlap, we are forced to extrapolate
one of the data sets to plot the light curve in a given energy band.
To derive the composite light curve
unambiguously for a prompt
 and an early X-ray emission, we need an observation 
of a prompt emission by current instruments,  
which overlap the energy range of XRT.

The tail behavior with $\alpha\lesssim2$ does not appear in
the uniform- and the Gaussian jet models; hence, it is important 
to constrain the jet structure.
However, in practical observations,
such gradually decaying prompt tail emission might be 
misidentified with the external shock component, as expected 
in the pre-{\it Swift} era.
Actually, some events have shown such a gradual decay, 
without the steep and the shallow decay phases, and their
temporal and spectral indices are consistent with
a classical afterglow interpretation \citep{obrien06b}.
Hence, in order to distinguish the prompt tail emission from 
the external shock component at a time interval $[T_a,T_b]$, 
one should study the spectral evolution and/or
the continuity and smoothness of the light curve 
(i.e., whether breaks appear or not) over the entire burst emission.

In this paper, we adopt $q_a=1\times10^{-2}$ and $q_b=1\times10^{-3}$ 
when the fitting epoch $[T_a,T_b]$ is determined [see Eq.~(\ref{eq:Tab})].
Then, the prompt tail emission in this time interval is so dim that
it may often be obscured by the external shock component, causing a
subsequent shallow decay phase of the X-ray afterglow.
One possible way to resolve this problem is to adopt larger
values of $q_{a}$ and $q_{b}$, 
e.g., $q_a=1/30$ and $q_b=1\times10^{-2}$,
in which the interval $[T_a,T_b]$ shifts toward earlier epochs,
so that the flux then is almost always dominated by the
prompt tail emission.
We have calculated the decay index distribution for this case
($q_a=1/30$ and $q_b=1\times10^{-2}$) and have found that the differences
between uniform- and power-law jets still arises as can be seen
in the case of $q_a=1\times10^{-2}$ and $q_b=1\times10^{-3}$,
 so that our conclusion remains unchanged.
However, the duration of the interval, $T_b-T_a$, becomes 
short, which might prevent us from observationally fixing
the decay index at high significance.
If $q_a\gtrsim1/30$, the emission at $[T_a,T_b]$ is dominated
by the last brightest pulse.  
Then, the light-curve shape at $[T_a,T_b]$
does not reflect the global jet structure, but reflects
the properties of the emitting subshell causing the last brightest pulse.
Another way to resolve the problem is to remove the shallow decay
component. For this purpose, the origin of the shallow decay phase
should be clarified in order to extract the dim prompt tail emission
exactly.
The other problem is contamination of  X-ray flares, whose
contribution has to be removed in order to investigate the tail 
emission component.
In any case, if the GRB occurs in an extremely low-density region 
(a so-called naked GRB),
where the external shock emission is expected to be undetectable,
our method may be a powerful tool to investigate the GRB jet structure.

%
\acknowledgements
This work was supported in part by Grants-in-Aid for Scientific Research
of the Japanese Ministry of Education, Culture, Sports, Science,
and Technology 18740153 (R.~Y.).
T.S. was supported by an appointment of the NASA Postdoctoral Program 
at the Goddard Space Flight Center, administered by Oak Ridge Associated 
Universities through a contract with NASA.

\clearpage

%
%


%

\end{document}